\documentclass{article}

\usepackage{epsfig}

\begin{document}

\title{Phenomenological approach to profile impact of scientific research: Citation
Mining\thanks{%
The views in this paper are solely those of the authors, and do not
necessarily represent the views of the U.S. Department of the Navy or any of
its components, the Universidad Nacional Aut\'{o}noma de M\'{e}xico or
NOESIS Inc.}}
\author{J. A. del R\'{\i}o, \\
Centro de Investigaci\'on en Energ\'{\i}a, \\
Universidad Nacional Aut\'onoma de M\'exico, \\
A.P. 34 Temixco, Mor. Mexico \and R. N. Kostoff, \\
Office of Naval Research Arlington, VA, USA \and E. O. Garc\'{\i}a, A. M. 
Ram\'{\i}rez 
\\
Centro de Investigaci\'on en Energ\'{\i}a, \\
Universidad Nacional Aut\'onoma de M\'exico, \\
A.P.34 Temixco, Mor. M\'exico \and and J. A. Humenik \\
NOESIS, Inc. Manassas, VA, USA}
\maketitle

\begin{abstract}
In this paper we present a phenomenological approach to describe a complex
system: scientific research impact through Citation Mining. The novel
concept of Citation Mining, a combination of citation bibliometrics and text
mining, is used for the phenomenological description. Citation Mining starts
with a group of core papers whose impact is to be examined, retrieves the
papers that cite these core papers, and then analyzes the
technical infrestructure (authors, jorunals, institutions) 
of the citing papers as well as their 
thematic characteristics. The Science Citation Index is used as the source
database for the core and citing papers, since its citation-based structure
enables the capability to perform citation studies easily. This paper
presents illustrative examples in photovoltaics (applied research) and
sandpile dynamics (basic research) to show the types of output products
possible. Bibliometric profiling
is used to generate the technical infrastructure, and 
is performed over a number of the citing
papers' record fields to offer different perspectives on the citing (user)
community. Text mining is performed on the aggregate citing papers, to
identify aggregate citing community themes, and to identify extra-discipline
and applications themes. The photovoltaics applied research papers had on
the order of hundreds of citations in aggregate. All of the citing papers
ranged from applied research to applications, and their main themes were
fully aligned with those of the aggregate cited papers. This seems to be the
typical case with applied research. The sandpile dynamics basic research
papers had hundreds of citations in aggregate. Most of the citing papers
were also basic research whose main themes were aligned with those of the
cited paper. This seems to be the typical case with basic research. 
However, about
twenty percent of the citing papers were research or development in other
disciplines, or development within the same discipline. There was no-time
lag between publication and citation by the extra-discipline research papers
, but there
was a four year lag time between publication and citation by the development
papers.
\end{abstract}

\section{INTRODUCTION}

Most scientists publish their findings to disseminate their research results
widely and hope that their research has some impact in the scientists'
community and society in general. For many years, citation counts have been
used for this goal. This research evaluation approach has produced
interesting results identifying the complex nature of Physics impact in the
research community (for instance see refs. \cite{katz}, \cite{vanraan}, \cite
{kos2}). This identification of diverse research impacts is
important to research managers/ sponsors/ evaluators, and of course
performers. They are interested in the types of people and organizations
citing the research outputs, and whether the citing audience is the target
audience. Also, they are interested in whether the development categories
and technical disciplines impacted by the research outputs are the desired
targets. Since fundamental research can evolve along myriad paths, tracking
diverse impacts becomes very complex.

Recently, scientists have addressed the problem of citation in scientific
research from different perspectives: looking for topological description of
citation \cite{bilke}, or for power laws in citation networks \cite{redner},
or obtaining power laws in number of cites received by journals according to
their number of published papers \cite{katz}, or though two-step competition
model relating the number of publications and number of cites. These
different approaches use power laws trying to obtain simple results from
complex interactions, assuming that the precise details of the interactions
among the parts of the system play no role in determining the overall
behavior of the system\cite{stanley}. Other approaches try to find some kind
of universality in the behavior of research institutions \cite{amaral},\cite
{nature}. However, in order to obtain a detailed representation of the
system it is important to know the details of the interactions. 
The creation of roadmaps for science and technology illuminates these 
interactions, and allows the progression of research to be portrayed
from both retrospective and prospective perspectives. 
and with these
\cite{kos01}, \cite{Galvin}. 
The analysis of the cites and authorship from a social network
perspective gives other useful information to characterize scientific
disciplines \cite{liberman}, \cite{newman}. However, these approaches:
scaling, networks and roadmaps, have limitations due to the
fact that they explore only partially the data available from the citation 
system.
The detailed analysis of all the available data of the citing community is
required to obtain more information and knowledge. Until now there has been no
comprehensive systematic methodology to deal with the information available 
through
cites of the scientific article. To overcome the above mentioned limitations
of these techniques, we have developed a phenomenological approach 
to deal with all the citation
information available, and obtain a more detailed description of this complex
system. The aim of this paper is to show how we can obtain a more
complete profile 
of the citing papers, and thereby get a more
complete representation of the impact of science. The application of this
kind of phenomenological detailed description is useful to obtain a
different and illustrative view of the complex systems.

The enhanced coverage of the research literature by the Web version of the
Science Citation Index (SCI-\symbol{126}5300 leading research journals)
allows a broad variety of bibliometric analyses of R\&D units (papers,
researchers, journals, institutions, countries, technical areas) to be
performed.

Aggregation of citation number counts is characteristic of almost all
published citation studies\cite{katz},\cite{redner}, \cite{amaral}; this
approach identifies R\&D units that have had (and have not had) gross impact on
the user community. However, as we have already mentioned, this 
abscence of fine-structure represents a limited perspective; therefore, 
we require an approach
that utilizes all the available citing data and could help answer questions
such as:

What types of people and organizations are citing the research outputs; is
this the desired target audience?

What development categories are citing the research outputs?

What technical disciplines are citing the research outputs?

What are the relationships between the citing technical disciplines and the
cited technical disciplines?

The aim of the present study is to show the power and capability of this new
phenomenological approach to citer profiling. It is necessary to stress that
the aim is not to assess the productivity and magnitude of impact of any
individual researcher, research group, laboratory, institution, or country.
To perform such an assessment, the authors would need a charter and
statistically representative data based on the unit of assessment, i.e. to
make a portrait of the research impact according with the current available
scientific databases.

The organization of this paper is the following: In the second section, we
summarize the methodology, and describe citation mining. In the third
section, we present the results of analyzing four sets of papers. In the
following we will refer to these papers as cited papers, in order to
distinguish them from the citing papers that are the papers in which the
cited papers are referenced. We profile the four selected paper sets through
the analysis of the characteristics of the citing authors, journals where
the citing papers appeared, references in the citing papers, and also
through the analysis of linguistic correlations in the abstract of the cited
and citing papers as well as the titles and the keywords, and other registers
available in the SCI database. Finally, we conclude with some remarks on the
present study.

\section{METHOD}

In this section we describe a simple procedure to incorporate much of the
SCI information in the analysis of the impact of scientific papers. First,
we identify the types of data contained in the SCI (circa early 2000), and
the types of analyses that will be performed on this information (see Table
1).

Table 1 shows a record from the SCI, without the field tags. The actual
paper that it represents is referred in the following description as the
'full paper'. Starting from the top, the individual fields are:

\strut

\noindent 
\begin{tabular}{||p{1.5in}|p{2.5in}||}
\hline\hline
\textbf{Table 1} & \textbf{SCI RECORD} \\ \hline
Title & the complete title of the actual paper \\ \hline
Authors & all the authors of the actual paper \\ \hline
Source & journal name (e.g. Nature) \\ \hline
Issue/ Page(s)/ Publication Date &  \\ \hline
Document Type & Article, notes, review, letters \\ \hline
Cited References & the number and names of the references cited in the full
paper. \\ \hline
Times Cited & the number and names of the papers (whose records are
contained in the SCI) that cited the full paper. Thus, the number provided
by this field is a lower bound. \\ \hline
Related Records & papers sharing at least one reference with the SCI 
record \\ \hline
Abstract & the complete Abstract from the full paper \\ \hline
Keywords Plus & keywords supplied by the indexer. In this example, no
Keywords were supplied by the author, but the SCI contains a field for
Keywords, if supplied. \\ \hline
Addresses & organizational and street addresses of the authors. For multiple
authors, this can be a difficult field to interpret accurately. Different
authors from the same organizational unit may describe their organizational
level differently. Different authors may abbreviate the same organizational
unit differently. \\ \hline
Publisher &  \\ \hline\hline
\end{tabular}

\strut

Before proceeding further, we define the overall study objectives, followed
by the approach chosen. The purpose is to identify the infrastructure and
technical characteristics of the citing community, by stratified categories.
The approach is to perform bibliometric analysis of the citing papers to
identify the infrastructure, text mining (extraction of useful information
from text using computational linguistics mainly) of the citing papers to
identify the main technical themes and their relationships, and then
integrate the bibliometrics and text mining results to obtain a unified
picture of the citing community.

Specifically, we want to present a summary analysis of the citing paper
information. This establishes boundaries on the population to be cited, the
fields of the cited records to be analyzed, and the statistical requirements
for the citing population. The population to be cited could be (and is) an
individual paper, the papers from a single author, the papers from an
organizational unit, the papers from a technical discipline, the papers from
a country, and/ or various combinations of the above. The key fields of the
citing records will be analyzed.

Title record is used in text mining together with the other unstructured
text fields, Abstracts and Keywords, to perform the correlation analysis of
the themes in the cited paper to those of the citing papers. Computational
linguistics analysis is then performed.

Author records used to obtain multi-author distribution profiles could be
computed (e.g., number of papers with one author, number with two authors,
etc.).

Counts in Source field can lead to journal name distributions, theme
distributions, and development level distributions.

Document Type register allows distributions of different document types to
be computed (e.g., three articles, four conference proceedings, etc.)

Language field allows distributions over languages to be computed.

Cited References allows a historical analysis of the problem to be
performed, and this field can be used to analyze the interrelations between
different groups working on related problems.

Times Cited register would be important if the citing papers are of
sufficient vintage. Then their multiplier effect would be of interest, and
could be computed. The distribution profile of times cited of the citing
paper would be generated.

The Addresses register allows distributions of names and types of
institutions, and countries, to be generated. Institution and country
combinations would be of special interest, and could be correlated with
author combination distributions.

The present demonstration of citation mining includes a comparison of a
cited research unit from a developing country with a cited research unit
from a developed country. It also compares a cited unit from a basic
research field with a cited unit from an applied research field.
Specifically, the technique is being demonstrated using selected papers from
a Mexican semiconductor applied research group (MexA), a United States
semiconductor applied research group (USA), a British fundamental research
group (BriF), and a United States fundamental research group (USF) (see
Table 2). These papers were selected based on the authors' familiarity with
the topical matter, and the desire to examine papers that are reasonably
cited. Here, we select these analyzed sets considering at least 50 external
cites in order to have a good phenomenological description.

\strut

\noindent 
\begin{tabular}{||l|l|l||}
\hline\hline
\textbf{Table 2} &  & \textbf{Cited Papers Used for Study} \\ \hline
Group & Times Cited & PAPERS \\ \hline
MexA & 59 & 
\begin{tabular}{l}
Nair, 1988 Sem. Sc. Tech. 3 , 134 \\ 
Nair, 1989 J Phys D - Appl Phys 22, 829 \\ 
Nair, 1989 Sem. Sc. and Tech. 4, 191 \\ 
Nair, 1994 J Appl Phys, 75 , 1557
\end{tabular}
\\ \hline
USA & 88 & 
\begin{tabular}{l}
Tuttle, 1995, Prog. Photovoltaic 3, 235 \\ 
Gabor, 1994, Appl. Phys. Lett. 65, 198 \\ 
Tuttle, 1995, J. Appl. Phys. 78, 269 \\ 
Tuttle, 1995, J. Appl. Phys. 77, 153 \\ 
Nelson, 1993, J. Appl. Phys. 74, 5757
\end{tabular}
\\ \hline
BriF & 119 & 
\begin{tabular}{l}
Mehta, 1989, Physica A, 157, 1091 \\ 
Mehta, 1991, Phys Rev Lett, 67, 394 \\ 
Barker, 1992, Phys Rev A, 45, 3435 \\ 
Mehta, 1996, Phys Rev E, 53, 92
\end{tabular}
\\ \hline
USF & 307 & Jaeger, 1992, Science, 255, 1523 \\ \hline\hline
\end{tabular}

\strut

In addition, selection and banding of variables are key aspects of the
bibliometric study. While specific variable values are of interest in some
cases (e.g., names of specific citing institutions), there tends to be
substantial value in meta-level groupings (e.g., institution class, such as
government, industry, academia). Objectives of the study are to demonstrate
important variables, types of meta-level groupings providing the most
information and insight, and those conditions under which
non-dimensionalization become useful. However, we present also two analyses
at the micro-level involving specific correlations between both citing
author and references for BriF and USF papers. This latter analysis is
directly important for the performers of scientific research. In addition,
text mining could be performed on the text fields (mainly the Abstract, but
including the Title and Keywords) to supplement the analysis on the
semi-structured and structured fields.

\section{RESULTS}

This section presents the bibliometric and text mining results, showing the
advantages and broad perspectives offered by these techniques, both alone
and combined. The results are presented in graph and tabular forms. In order
to organize the presentation, we divide our results into bibliometrics and
textmining analysis.

\subsection{Bibliometrics Analysis}

Figure 1 contains a bar graph of multi-author distribution for the four sets
analyzed. The ordinate represents the fraction of total papers published in
each author band, and the abscissa represents the number of authors per
paper. The most striking feature of this graph is the behavior at the wings.
The papers citing basic research dominate the low end (single author), while
the papers citing applied research dominate the high end (6-7 authors). The
papers citing basic research (BriF and USF) have a similar number of authors
per paper, with a maximum in the frequency distribution at two authors per
paper. The USA citing papers show gaussian-like authorship distribution with
three and four authors per paper, while the MexA group citing papers show a
distribution similar to the groups citing fundamental research papers but
with fewer single-author papers. These four sets show author distributions
where 90\% of the papers had less than six authors. These results confirm
the diversity of collaborative group compositions over different disciplines
and levels of development.

Generally, as projects become more applied, they tend to become larger and
more expensive, and require more resources. They also usually require the
integration of multiple disciplines. Both these characteristics typically
result in larger research groups, and hence in more contributors to a
project and its resulting documents. Experimental work usually involves
larger teams than theoretical work, while modeling and simulation activities
tend to allow more individual efforts. The strong experimental emphasis of
the two applied semiconductor groups, with little evidence of computer
simulation shown, results in large teams on average. The more balanced
theory/ experiment combination of the basic research group tends to suppress
larger team efforts in favor of more individualized research. In addition,
the intrinsic nature of sandpile vibration research, as opposed to
elementary particle or fusion research, does not require large facilities
and large research teams.

The citing journal discipline frequency is shown in Table 3. Clearly, each
paper set has defined its main discipline well. Also, there is a symmetry in
the cross citing disciplines. USF and BriF groups were cited around 80\% in
fundamental journals (Phys, Bio, Chem) and close to 10\% in applied journals
(Environ, Mate, Eng.). Similarly, MexA
and USA groups were cited close to 50\% in applied journals and 45\% in
fundamental journals. These journal discipline results suggest that the
applications developed by the MexA group have a stronger impact on chemical
journals than those of the USA group, while the applications developed by 
the USA have a group stronger impact on
physics journals than those of the MexA group. A point to be stressed is 
that only the fundamental papers
received cites in journals clearly outside of their disciplines.

\begin{tabular}{||l|l|l|l|l||}
\hline\hline
\textbf{Table 3} &  & \textbf{Citing } & \textbf{Journal } & \textbf{Theme}
\\ \hline
\textit{Theme} & \textit{MexA} & \textit{USA} & \textit{BriF} & \textit{USF}
\\ \hline
Phys. & 20\% & 44\% & 83\% & 79\% \\ \hline
Bio. & 0 & 0 & 0 & 1\% \\ \hline
Chem. & 25\% & 1\% & 2\% & 2\% \\ \hline
Enviromental & 0 & 0 & 1\% & 0 \\ \hline
Mate. \& Eng. & 54\% & 55\% & 12\% & 13\% \\ \hline
Multidiscipl. & 0 & 0 & 2\% & 5\% \\ \hline\hline
\end{tabular}

The discipline distribution of the citing papers, produced by analyzing the
papers' Abstracts and Titles, is shown in Table 4. It is slightly different
from Table 3. As concluded in the text mining, these free-text fields
provide far more precise information than can be obtained from the journal
discipline. Multi-disciplinary journals can publish uni-disciplinary papers
from many different disciplines. Also, the journal categories, determined by
ISI, are not a unique reflection of specific contents (e.g., an
environmental journal can accept engineering papers, a materials journal can
accept physics papers, etc.). 

\begin{tabular}{||l|l|l|l|l||}
\hline\hline
\textbf{Table 4} &  & \textbf{Citing } & \textbf{Article } & \textbf{Theme}
\\ \hline
\textit{Theme} & \textit{MexA} & \textit{USA} & \textit{BriF} & \textit{USF}
\\ \hline
Phys. & 10\% & 22\% & 86\% & 83\% \\ \hline
Bio. & 0 & 0 & 0 & 1\% \\ \hline
Chem. & 26\% & 5\% & 0\% & 1\% \\ \hline
Mate. \& Eng. & 64\% & 73\% & 12\% & 14\% \\ \hline
Multidiscipl. & 0 & 0 & 2\% & 1\% \\ \hline\hline
\end{tabular}

In three of the four sets analyzed, the component papers were published in
different years (see Figure. 2). The MexA set was published from 1989 to
1994, USA from 1994 to 1995, BriF from 1989 to 1996, while USF includes only
one paper published in 1992. Figure 2 shows a clear oscillating behavior of
USA and BriF, due partly to the different dates of paper publication. Also,
most of the sets have between 10\% and 20\% of cites per year, while the USA
set received 38\% of the cites in 1998.

The single highly-cited paper feature of the USF set allows additional
analyses and perspectives. In Figure 3, the USF citing paper disciplines are
shown as a function of time. As time evolves, citing papers from disciplines
other than those of the cited paper emerge. An important point is the
four-year delay of the systematic appearance of application/development 
citing papers, but no delay for extra-discipline research citing papers.

Table 5 shows that most cites appear in articles. The four analyzed sets are
cited in review articles and letters. This indicates the relevance of the
analyzed papers. One important point is that only the fundamental papers are
cited in notes, and only the USF paper was cited in an editorial document.

\begin{tabular}{||l|l|l|l|l||}
\hline\hline
\textbf{Table 5} &  & \textbf{Citing } & \textbf{Paper } & \textbf{Type} \\ 
\hline
\textit{Paper type} & \textit{MexA} & \textit{USA} & \textit{BriF} & \textit{%
USF} \\ \hline
Article & 95\% & 96\% & 92\% & 89\% \\ \hline
Letter & 3\% & 1\% & 2\% & 2\% \\ \hline
Review & 2\% & 3\% & 2\% & 5\% \\ \hline
Note & 0 & 0 & 4\% & 3\% \\ \hline
Edit.Mat. & 0 & 0 & 0 & 1\% \\ \hline\hline
\end{tabular}

Table 6 shows that English is the dominant language of all the paper sets
analyzed. However, the surprising appearance of a significant number of
citing papers written in Romanian for the MexA set indicates that MexA's
work is important for at least one developing country.

\begin{tabular}{||l|l|l|l|l||}
\hline\hline
\textbf{Table 6} &  & \textbf{Citing } & \textbf{Paper } & \textbf{Language}
\\ \hline
\textit{Language} & \textit{MexA} & \textit{USA} & \textit{BriF} & \textit{%
USF} \\ \hline
English & 93\% & 100\% & 99\% & 99.7\% \\ \hline
Romanian & 7\% & 0 & 0 & 0 \\ \hline
French & 0 & 0 & 1\% & 0 \\ \hline
German & 0 & 0 & 0 & 0.3\% \\ \hline\hline
\end{tabular}

Figure 4 shows the profile of the citing institutions. Clearly, academia has
the highest citing rates. Industry references the advance of
high-technological developments, but is not referencing directly the advances
in fundamental research. Research Centers follow applied and fundamental
research about equally. Direct government participation is not significant
in the fields studied. Government/ national laboratories were classified
under research centers.

There are 44 countries represented in the citing paper sets analyzed. Figure
5 shows only those countries with at least 10\% of the citing countries for
a set. USA has the most cites in aggregate. India has the largest cites of
the MexA set; Japan has the largest cites of the USA set. This fact is due
to the different nature of the applied technology developed by MexA and USA.
The USA set contains work related to high technology (high efficiency 
photovoltaic cells), and the MexA set is
dedicated to explore low-cost technology (low cost photovoltaic thin films).
 Therefore, this last set is cited
by the less affluent countries of India, Romania and Mexico. India and
Mexico also cite fundamental research, but not Romania. It is important to
stress that if no low-cost technology papers were considered, these latter
countries would not appear in this graph, and only developed countries would
appear. Another point is that England does not cite USA works.

Figure 6 shows clearly that the low-cost technology papers are cited by
developing countries. Developed countries cite the mostly high-technology
papers, there exist a clear asymmetry in the interests and of course the
number of cites from developing countries is less than cites from
developed ones.

The analysis of the most common citing authors is presented in Figures 7
where the frequency of an author citing USF (triangle) or BriF (square) is
plotted. Figure 7 shows that there is a closed relation between the citing
authors for both BriF and USF groups, actually there is a common citing
author who occupied the highest position in the frequency plot in both sets
(Hermann, HJ). Three of the highest citing authors are not shared between
the citing sets of USF and BriF. Jaeger and Nagel are the authors of the USF
paper and Mehta is one of the authors of BriF paper. They maintain awareness
of each other's work. In contradistinction, Figure 8 shows that MexA and USA
have no intersection between their topics (low cost photovoltaic thin films
and high efficiency photovoltaic cells, respectively), from the perspective
of the highest citing authors. Previous citation results have shown that
applied research authors tend to cite more fundamental research, along
relatively stratified lines. In Figure 8, it is clear that the maximum
citing author of the MexA group is a Romanian researcher.

In Figure 9, it is clear that there are common features in the number of
references in those papers that cite the core applied and fundamental
papers, but there are also some differences. For instance, at the lower end
of the spectrum (0-20), the applied papers' citing papers dominate. At the
higher end of the spectrum (21-50+), the fundamental papers' citing papers
dominate, with the exception of the BriF anomaly at 41-50.

There are many possible reasons for these differences, and separating out
the effects is complex. There are two different technical disciplines, and
each one has its citing culture and traditions. Also, each technical
discipline has a different level of research activity, and this could
influence the magnitude of citations generated. Basic researchers tend to
document more, and therefore produce a larger literature to cite. Finally,
there may be different citing practices in basic and applied research.

Frequency analysis of the most common references in the citing papers
provides insight to co-cited papers, and allows a historical perspective to
be obtained. The reference-frequency for the USF and BriF citing papers is
shown in Figure 10. Here we see that the fundamental papers dealing with
sand-piles are actually correlated, because the highest occurring
references in their citing papers are common references (each line has two
symbols).

In Figure 10 also we observed that Faraday's work (1831) appears within the
twenty papers most cited in the USF and BriF citing papers. This indicates
the fundamental and seminal character of the experimental work performed by
Faraday. Also, Reynolds' work (1885) appears within the twenty most cited
papers in the references of the BriF set. These two references also indicate
the longevity of the unsolved problems tackled by the USF and BriF groups.
The highest frequency co-cited papers have three interesting
characteristics. They are essentially all in the same general physics area,
they are all published in fundamental science journals (mainly physics), and
they are all relatively recent, indicating a dynamic research area with high
turnover.

The corresponding analysis of the most common references in the applied MexA
and USA groups is presented in Figure 11. These
two groups have no correlations, because they have no common references
between the highest occurring references in their citing papers. 
However, in the detailed analysis of the correlation there is
one paper in the intersection of these two groups.

This ends the bibliometric analysis. The following section illustrates the
usefulness of text mining analysis.

\subsection{Text mining}

The purpose of the text mining is to perform trans-citation linguistic
pattern analyses, and make trans-citation comparisons. Two text mining
techniques will be used for the following analyses. Phrase frequency
analysis will be used to identify the main technical themes of the citing
papers relative to the cited papers. Phrase proximity analysis, mainly
phrase clustering and taxonomy generation, will be used to show the
relationships among themes and category structures of the overall technical
citing disciplines. The findings for each of the four paper groups are
summarized. Phrase frequency results are presented for the first three
groups, and, in addition, phrase clustering results are presented for the
fourth group.

\subsubsection{USA}

The highest frequency single, adjacent double, and adjacent triple word
phrases from its USA citing papers aligned with the themes of the cited
paper can be seen in del Rio et al. \cite{delrio}.

The central themes and specific phrases used in the cited paper are
replicated in the citing papers. There were no phrases in the citing papers
that represented themes or disciplines significantly different from those in
the cited paper, above a frequency of unity. While there could possibly be
phrases representative of different themes with a unity frequency, some
minimal theme coherence was desired. The citing readership appears to be
strongly concentrated in the thematic areas of the cited paper. One suspects
that the audience obtained is the target audience for this paper, at least
in terms of thematic interest. We could not find phrases reflecting themes
other than cited paper.

\subsubsection{MexA}

The central themes and specific phrases used in the cited paper are
replicated in the citing papers. There were two phrases in the citing papers
that represented themes or disciplines other than those in the cited paper,
above a frequency of unity (See Table 7). These additional themes reflected
use of the solar coatings for automobile windows, in addition to the core
architectural (building) applications. This is a very small extrapolation.
Again, the citing readership appears to be strongly concentrated in the
thematic areas of the cited paper.

\strut

\noindent 
\begin{tabular}{||l|l||}
\hline\hline
\textbf{Table 7} & \textbf{Phrases Reflecting Themes other than Cited Paper
(MexA)} \\ \hline
Frequency & Theme \\ \hline
2 & AUTOMOBILE \\ \hline
2 & ARCHITECTURAL AND AUTOMOBILE \\ \hline\hline
\end{tabular}

\strut

\subsubsection{BriF}

Use of text mining capabilities, such as computational linguistics, allows
only those applications and extra-discipline papers of interest to be
identified, and the requisite information can then be obtained from reading
the Abstracts. In addition, the computational linguistics provides a
structure and categorization of these myriad applications, allowing the
larger context of application themes to be displayed and understood. While
phrase frequency algorithms were used for the present study, and proved
adequate, specifically-tailored co-occurrence and clustering algorithms are
being developed to improve the efficiency of the application papers
identification and retrieval process, see Table 8.

\strut

\noindent 
\begin{tabular}{||l|l||}
\hline\hline
\textbf{Table 8} & \textbf{Phrases Reflecting Themes other than Cited Paper
(BriF)} \\ \hline
Frequency & Theme \\ \hline
5 & CAR* \\ \hline
4 & ALLOY \\ \hline
4 & TRAFFIC \\ \hline
4 & WATER PROTON \\ \hline
3 & MAGNETIC \\ \hline
3 & PROTON TRANSVERSE \\ \hline
3 & TRAFFIC FLOW \\ \hline
2 & AQUIFER \\ \hline
2 & EXPRESSWAY \\ \hline
2 & FOOD \\ \hline
2 & FOODSTUSFFS \\ \hline
2 & ICE \\ \hline
2 & NUCLEAR \\ \hline
2 & TUNGSTEN \\ \hline
2 & WATER PROTON TRANSVERSE \\ \hline\hline
\end{tabular}

\strut

\subsubsection{USF}

The citing papers to USF paper representing different categories of
development and different disciplines from those of the cited paper are
portrayed graphically in Figure 12, the axes are Category, Alignment and
Papers. The Category represents the level of development characterized by
the citing paper (1=basic research; 2=applied research; 3=advanced
development/ applications), and the alignment represents the degree of
similarity between the main themes of the citing and cited papers (1=strong
alignment; 2=partial alignment; 3=little alignment).
 There are three interesting features on Figure 12. First, the tail
of total annual citation counts is very long, and shows little sign of
abating, this is one characteristic feature of a seminal paper. Second, the
fraction of extra-discipline basic research citing papers to total citing
papers ranges from about 20-40\% annually, 
with no latency period evident.
This instant extra-disciplinary diffusion may have been due to the
combination of intrinsic broad-based applicability of the subject matter and
publication of the paper in a high-circulation science journal with very
broad-based readership. Third, there was a four-year latency period before
the higher development category citing papers began to emerge. One can see
that black dots (earlier cites) are completely in the category. This
correlates with the results from the bibliometrics component. The latency
could have been due to the information remaining in the basic research
journals, and not reaching the applications community, or the time that an
application needs to be developed is of the order of four years. Thus, the
basic science publication feature that may have contributed heavily to
extra-discipline citations may also have limited higher development category
citations for the latency period.

The present phenomenological approach of identifying impact themes through
text mining allows a much more detailed and informative picture of the
impact of research to be obtained compared to semi-automated journal
classification comparison approaches \cite{davidse}. It represents the
difference between stating that a ''Physics paper impacted Geology
research'' and a ''paper focused on sand-pile avalanches for surface
smoothing impacted analyses of steep hill-slope landslides''.

In the final data analysis, a taxonomy of the USF citing papers was
generated using phrase clustering. The Abstracts of all the USF citing
papers were converted to phrases and their frequencies of occurrence with
use of a Natural Language Processor contained in the TechOasis software
package. The 153 highest frequency technical content phrases
(expert-selected) were exported to a statistical clustering software package
(WINSTAT). Based on the relations among phrases generated by this
package, a taxonomy was generated by the authors. A particularly helpful
output for each clustering run was the dendogram, a tree-like diagram
showing the structural branches that define the clusters. Figure 13 is one
dendogram based on the 48 highest frequency phrases (for illustration
purposes only). The abscissa contains the phrases that are clustered. The
ordinate is a distance metric. The smaller the distance at which phrases, or
phrase groups, are clustered, the closer is the connection between the
phrases. 

Thus, samples of the phrases combined, near the right hand end of the 
graph include DISSIPATION, COLLISIONS and
ENERGY, and MIXTURES/
ALTERNATING LAYERS/ SMALL GRAINS/ LARGE GRAINS/ STRATIFICATION. In the middle
part of the graph VIBRATION and AMPLITUDE can be found. At some later time,
the VIBRATION-AMPLITUDE combination is grouped with the GRAVITY-GRANULAR
MEDIA combination to form the next hierarchical level grouping, and so on.

Many statistical agglomeration techniques for clustering were tested; 
the Average Neighbor method
appeared to provide reasonably consistent good results. Analyses were
performed of the numerous cluster options that were produced. The following
is one of the top-level cluster descriptions that represented the results of
the phrase and word lists clustering best, as well as the factor matrix
clustering from the TechOasis results (more clusters can be found in Kostoff
et al. \cite{kostoff01}).

The highest level categorization based on the highest frequency 153 phrases
produced three distinct clusters \cite{kostoff01}: Structure/ Properties,
Flow-Based Experiments, Modeling and Simulation. In the description of the
Structure and Properties cluster (right part of the dendogram Figure 13)
that follows, phrases that appeared within the clusters will be capitalized.

This cluster contained MIXTURES of LARGE GRAINS and SMALL GRAINS, with
STRATIFICATION along ALTERNATING LAYERS based on SIZE SEGREGATION and grain
SHAPE and GEOMETRICAL PROFILE. The MIXTURE forms a PILE with an ANGLE of
REPOSE. When the ANGLE of REPOSE is LARGER than a critical ANGLE, DYNAMICAL
PROCESSES produce AVALANCHES, resulting in SURFACE FLOW within THIN LAYERS.

\section{CONCLUSIONS}

In this paper, we have presented a phenomenological technique to analyze some
aspects of a complex system like the multi-path non-monotonic impact of
scientific research. The result of using citation mining (bibliometrics and
text mining) to analyze the impact of science, through the use of the
available information from the Web of Science ISI, allows the profiling of
the citing papers of a given paper, research group, scientific organization,
etc. We illustrated citation mining through the analysis of four research
groups. This analysis provided multiple facets and perspectives of the
myriad impacts of research. Citation mining offers insights that would not
emerge if only separate citing paper counts were used independently, as is
the prevalent use of citation analysis today. Moreover, by removing the need
to actually read thousands of abstracts through the use of text mining,
comprehensive assessments of research impact become feasible. One important
result from the basic research citation mining \cite{kostoff01} was that
impacts are possible in myriad fields and applications not envisioned by the
researchers. This reference also questioned whether fundamental sand-pile
research would receive funding from Tokamak, air traffic control, or
materials programs, even though sand-pile research could impact these or
many other types of applications, as shown in the paper. The reference
concluded that sponsorship of some unfettered research must be protected,
for the strategic long-term benefits on global technology and applications!

Acknowledgments. This paper has been partially supported by DGAPA-UNAM under
project IN103100.

\newpage

Figure Captions

Figure 1. Multi-author distribution for the four paper sets analyzed .

Figure 2. Time profile of citing papers.

Figure 3. Time profile of citing papers for USF. Here it is important to
stress a four year delay in the appearence of applied citing papers, but
no delay in the appearence of extradiscipline fundamental papers.

Figure 4. Institution profile of citing papers. Most of the research
related
to these two fields is perfomed in academic institutions.

Figure 5. Country profile of citing papers. The main objective in the MexA
research is to obtain a low cost technology, this marks the countries
interested in this kind of work.

Figure 6. Citing country development phase. In this graph we are plotting
the relative number of citing countries according to their development
phase.  
Each axis indicates the ratio between developed or developing 
citing countries and the total number of citing countries to each group.
Lines are  drawn in order to guide
the eye and group the different origins of the citing papers.
In the fundamental groups, 
the difference between the number of papers produced in developed and
developing countries is clear. USF, USA and BriF receive more cites from 
developed
countries than from developing ones. However, note that again the topic of 
low cost 
technology is
more interesting for developing countries.

Figure 7. Correlation between the citing authors in fundamental sets. One
line with two symbols (triangle and square) means close relation.

Figure 8. Correlation between the citing authors in applied groups. Here,
there is no close relation.

Figure 9. Citing paper reference distribution.

Figure 10. Reference-frequency for fundamental sets.One line with two
symbols (triangle and circle) means that this reference appears in both
(BriF and USF) sets.

Figure 11. Reference-frequency for applied sets. There is no close relation,
because there is no line with two symbols.

Figure 12. Cathegory and alignment in the citing papers of USF. Black
symbols represent to earlies cites, while empty symbols indicate recent
cites.

Figure 13. Dendogram: tree-like diagram defining clusters of realted words
in the citing papers of USF. Lesser distance means close relations within
phrases.

\begin{figure}
\begin{center}
\epsfig{width=10cm,file=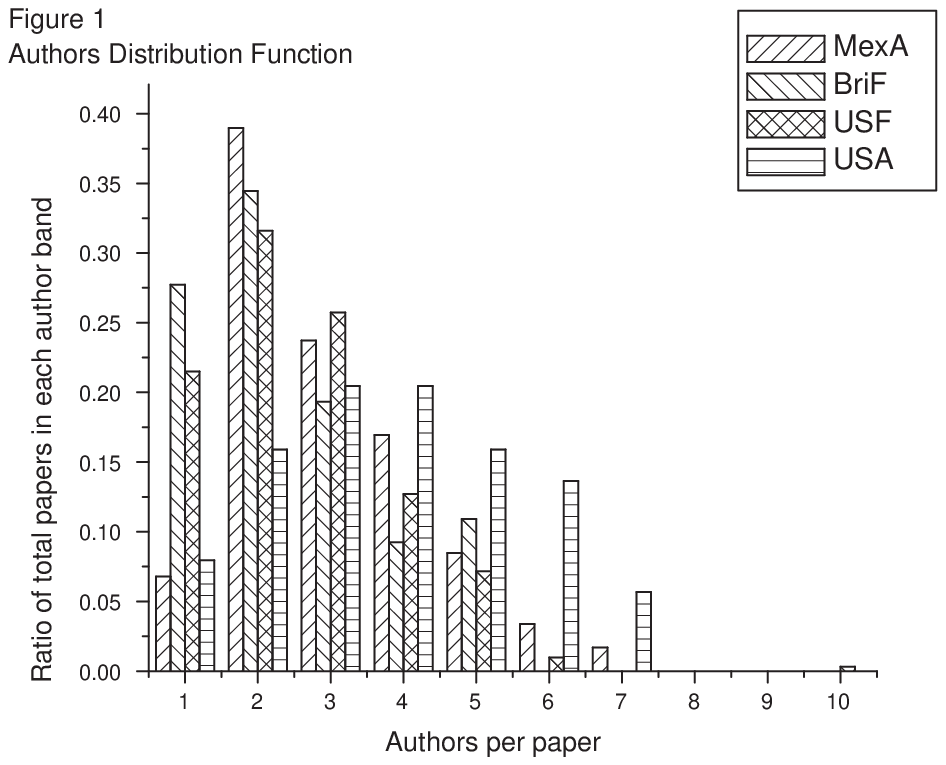}
\end{center}
\end{figure}

\begin{figure}
\begin{center}
\epsfig{width=10cm,file=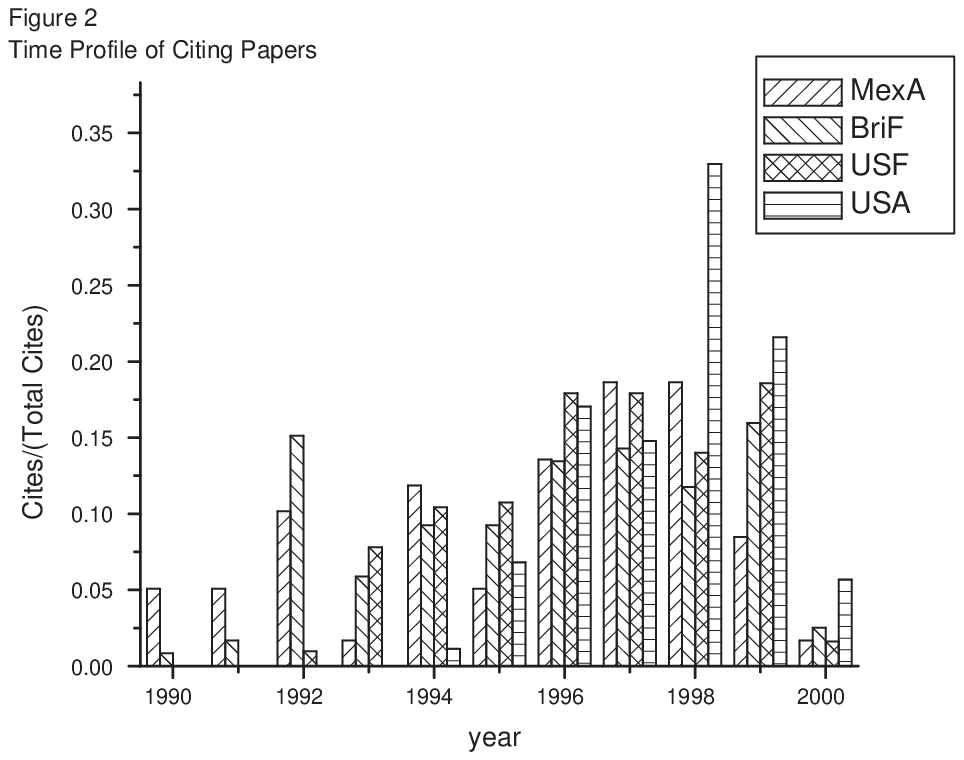}
\end{center}
\end{figure}

\begin{figure}
\begin{center}
\epsfig{width=10cm,file=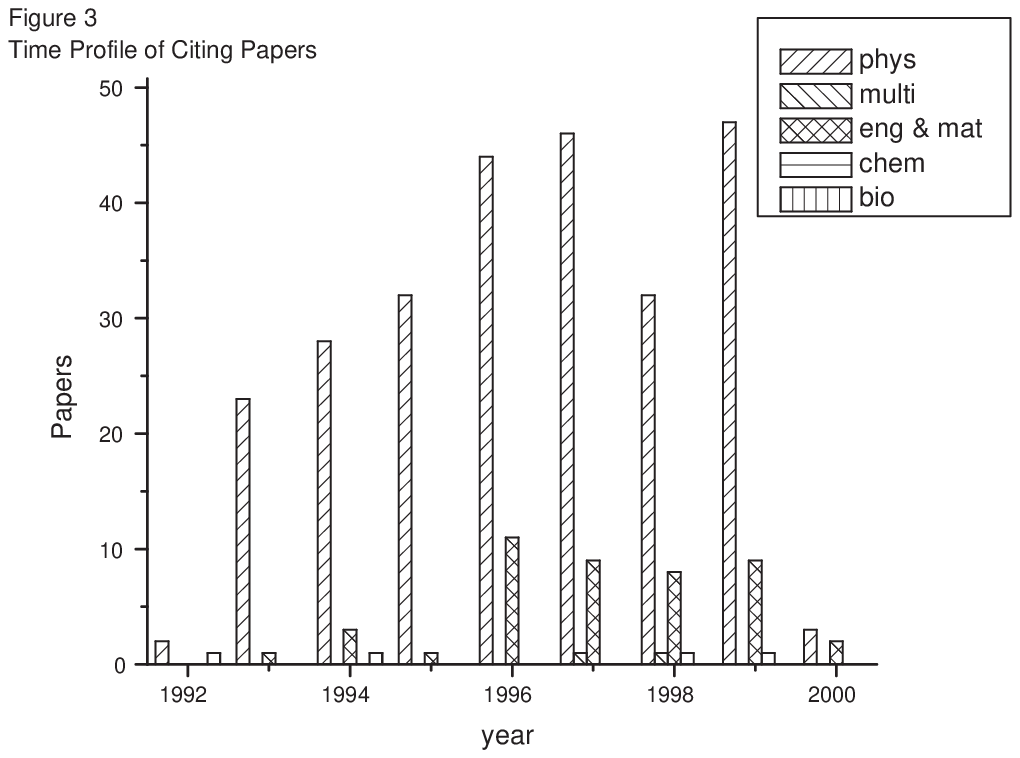}
\end{center}
\end{figure}

\begin{figure}
\begin{center}
\epsfig{width=10cm,file=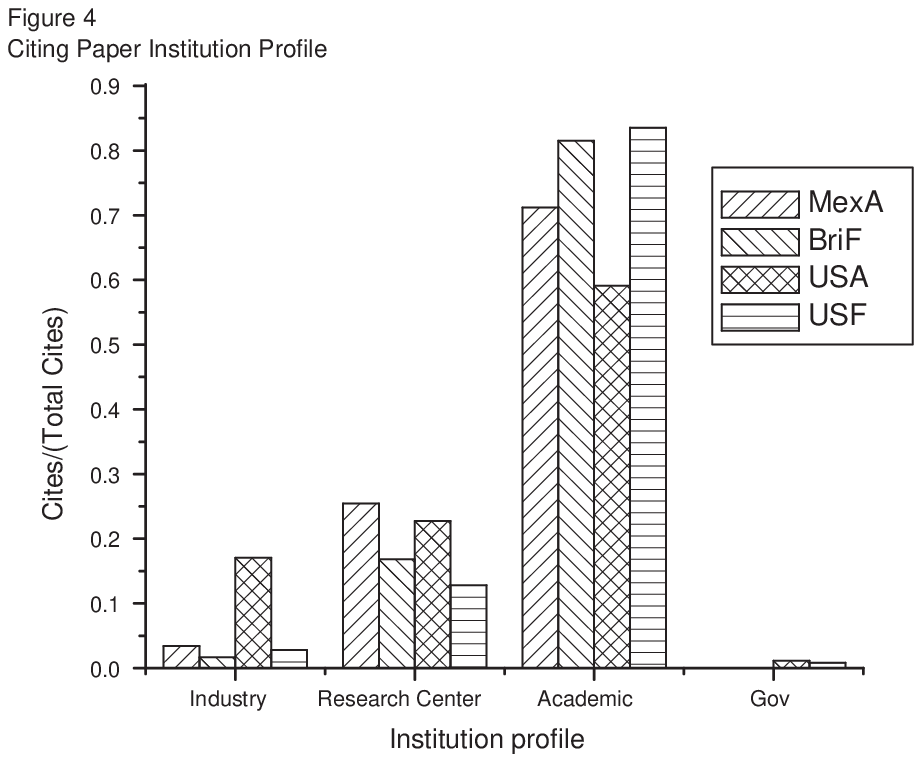}
\end{center}
\end{figure}

\begin{figure}
\begin{center}
\epsfig{width=10cm,file=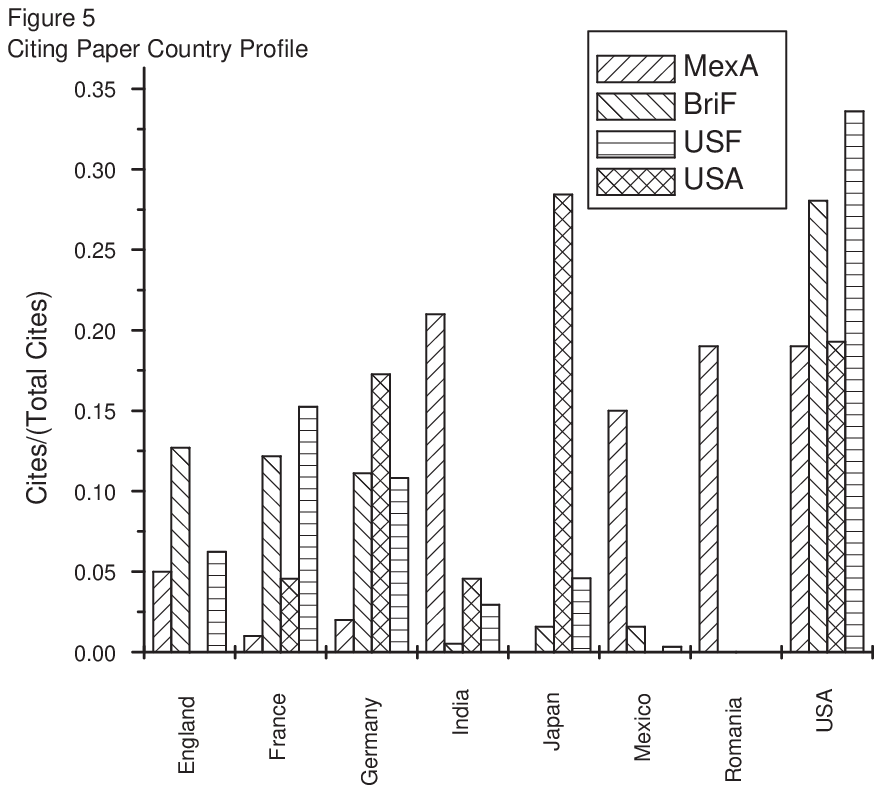}
\end{center}
\end{figure}

\begin{figure}
\begin{center}
\epsfig{width=10cm,file=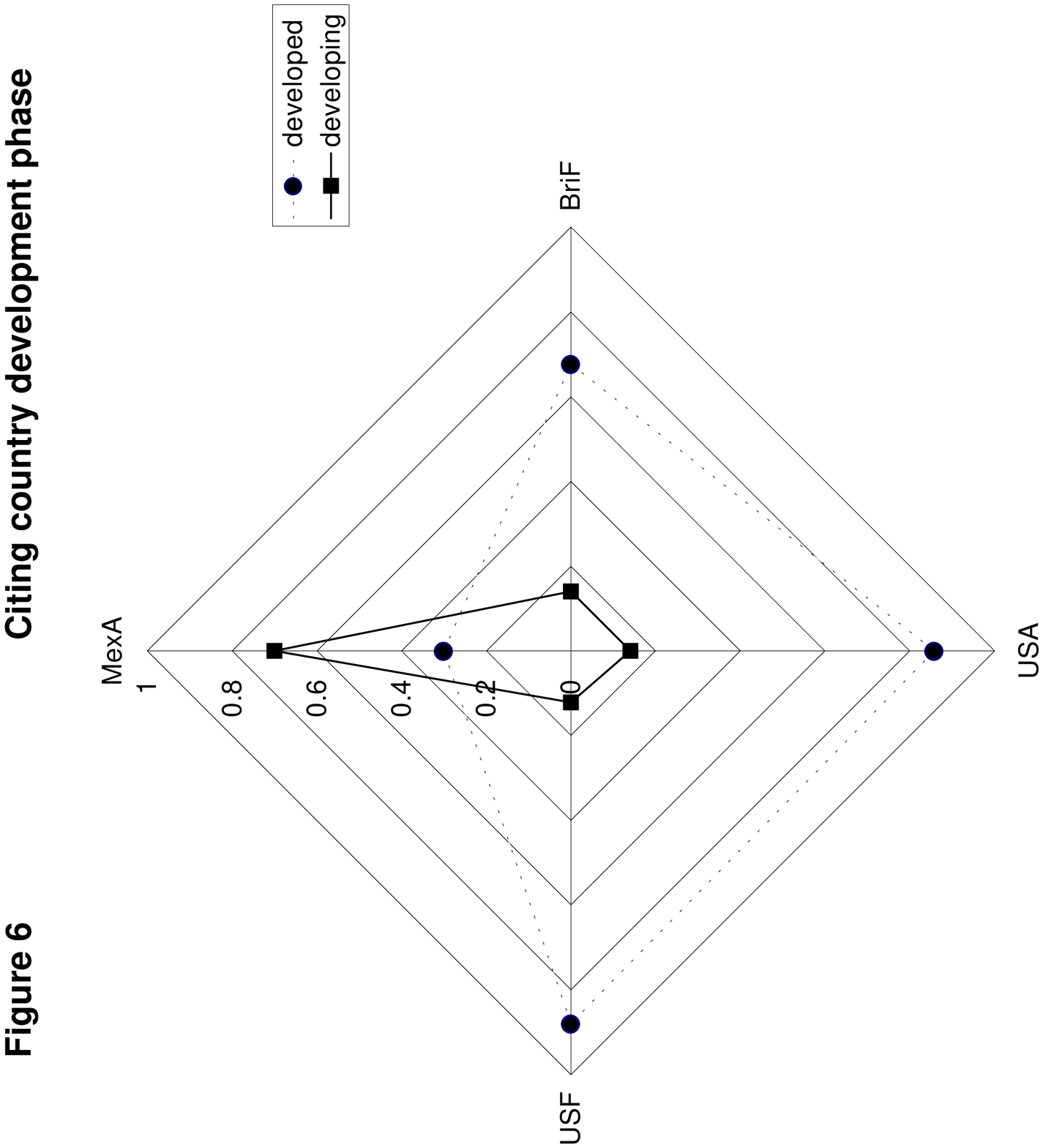}
\end{center}
\end{figure}

\begin{figure}
\begin{center}
\epsfig{width=10cm,file=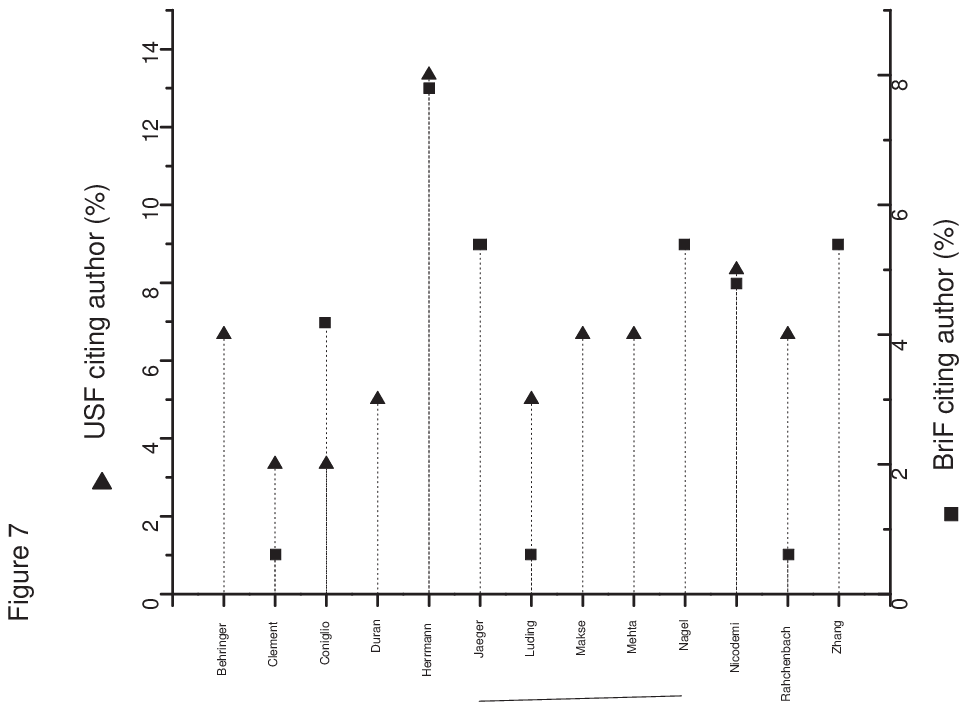}
\end{center}
\end{figure}

\begin{figure}
\begin{center}
\epsfig{width=10cm,file=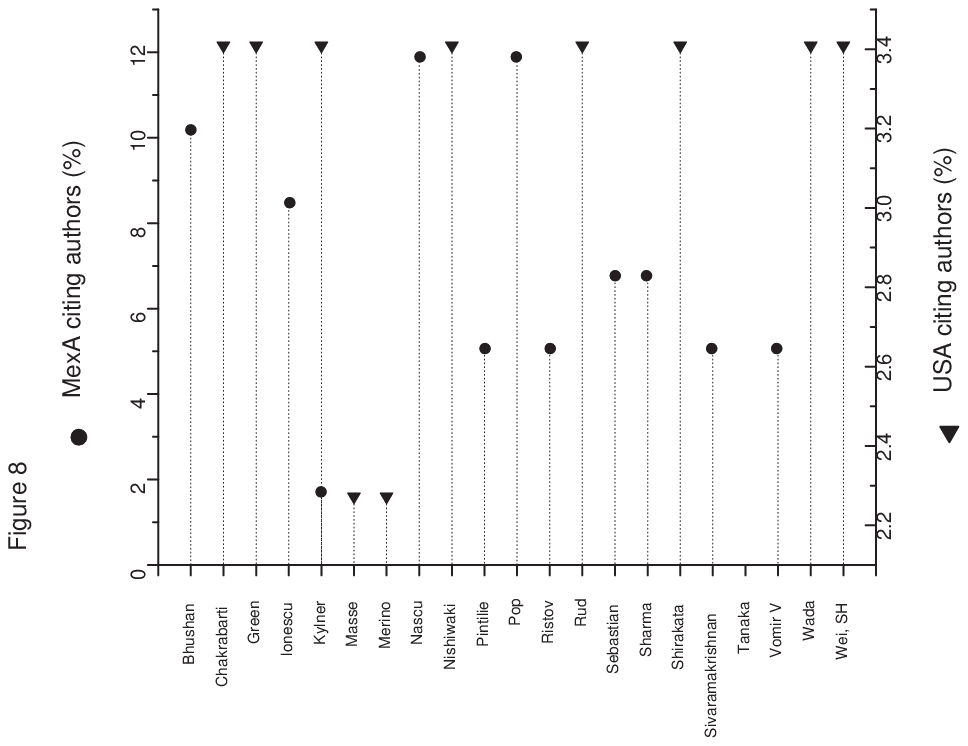}
\end{center}
\end{figure}

\begin{figure}
\begin{center}
\epsfig{width=10cm,file=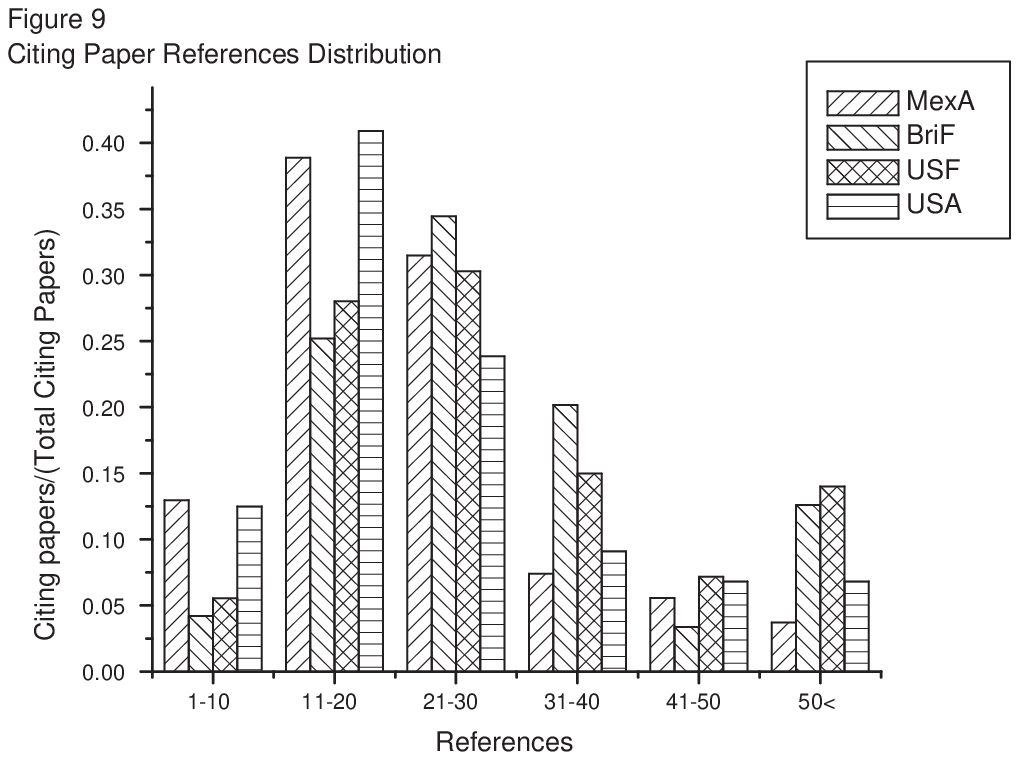}
\end{center}
\end{figure}

\begin{figure}
\begin{center}
\epsfig{width=10cm,file=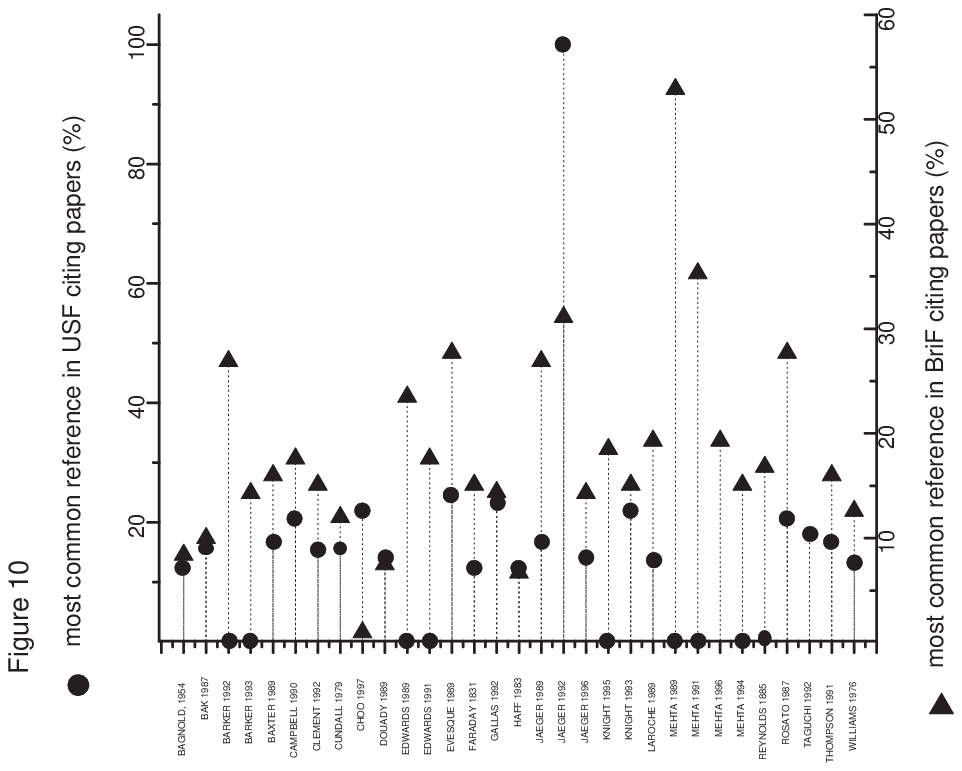}
\end{center}
\end{figure}

\begin{figure}
\begin{center}
\epsfig{width=10cm,file=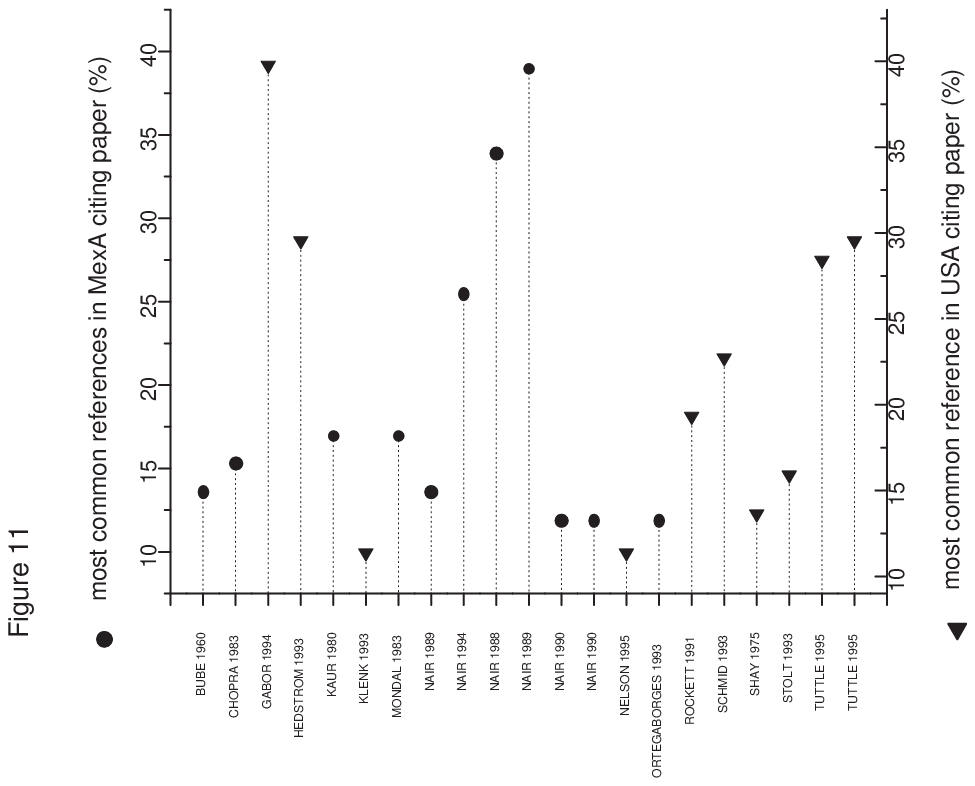}
\end{center}
\end{figure}

\begin{figure}
\begin{center}
\epsfig{width=10cm,file=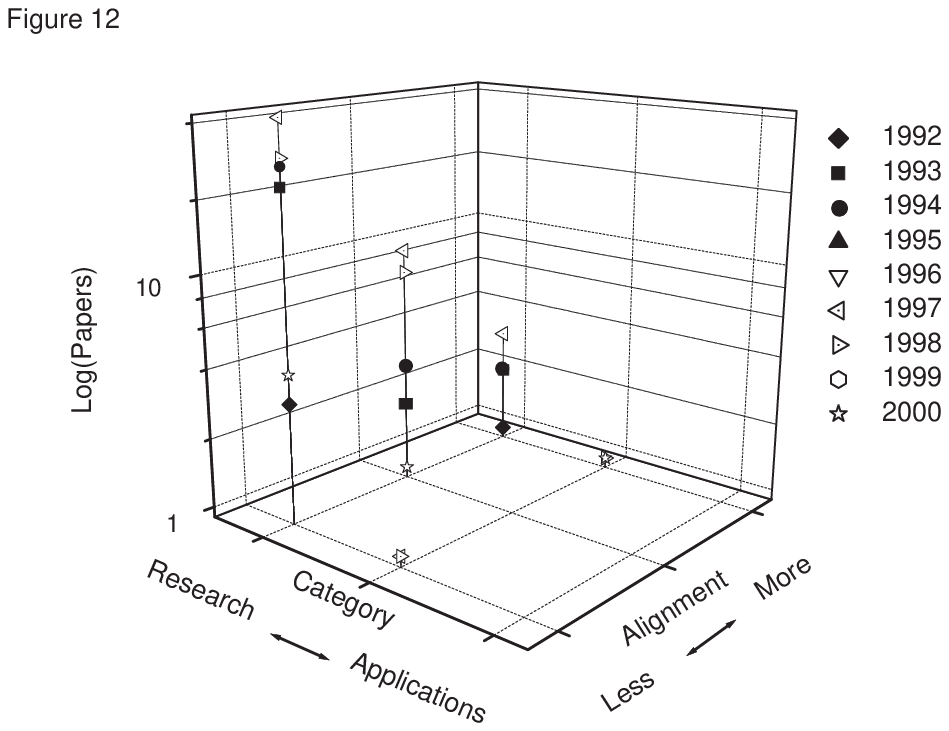}
\end{center}
\end{figure}

\begin{figure}
\begin{center}
\epsfig{width=10cm,file=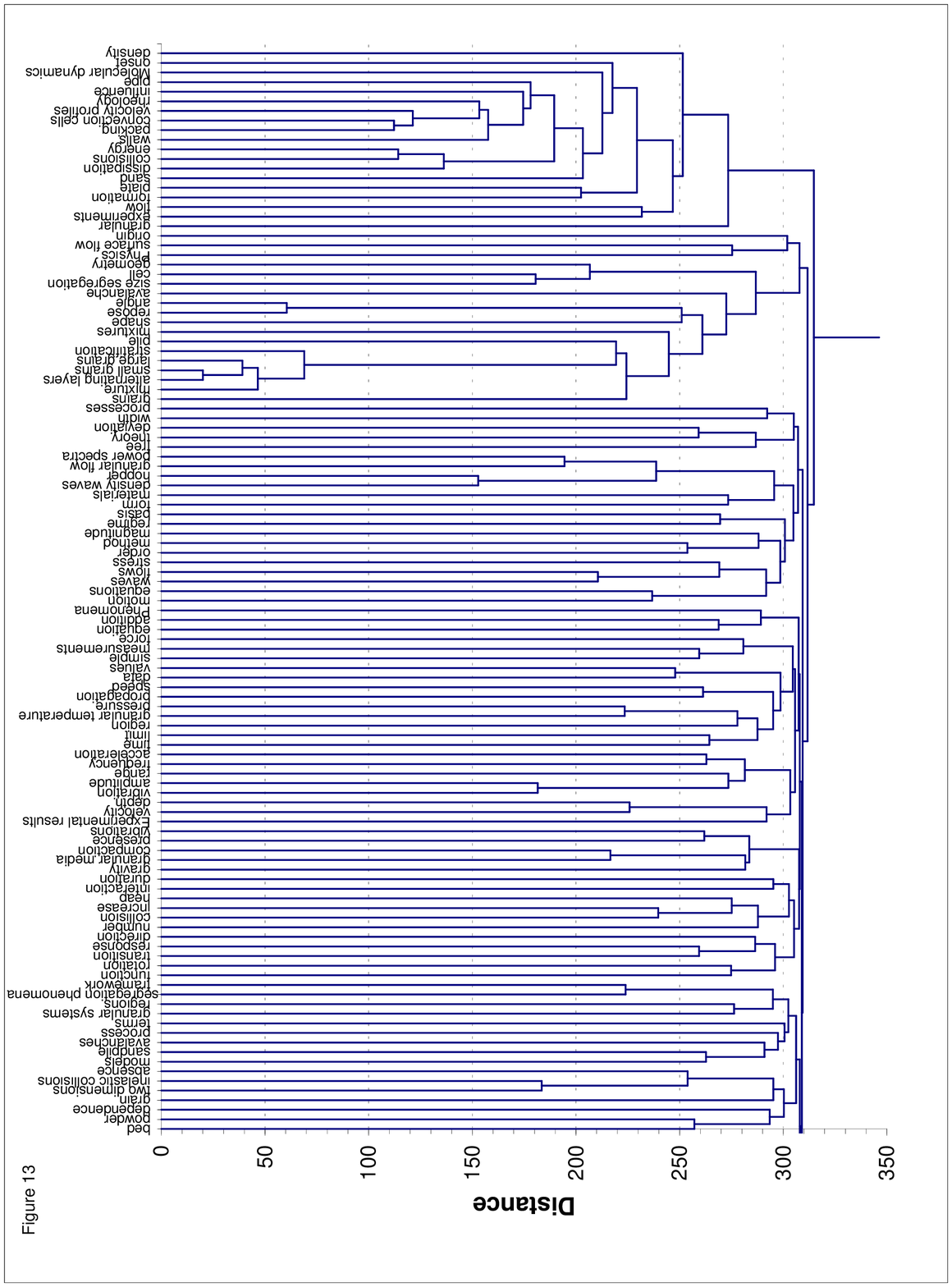}
\end{center}
\end{figure}

\end{document}